\documentclass[graybox]{svmult}

\usepackage{mathptmx}       
\usepackage{helvet}         
\usepackage{courier}        
\usepackage{type1cm}        
\usepackage{makeidx}         
\usepackage{graphicx}        
\usepackage{multicol}        
\usepackage[bottom]{footmisc}

\newcommand{\red}[1]{{}}
\newcommand{\blue}[1]{{#1}}

\newcommand{\bd}[1]{\mathbf{#1}}

\newcommand{\x}{{\bd{x}}}

\newcommand{\p}{{\bd{p}}}

\renewcommand{\r}{{\bd{r}}}

\makeindex             

\begin{document}

\title*{A Review of Wave Packet Molecular Dynamics}
\author{Paul E. Grabowski} 


\institute{Paul E. Grabowski \at University of California, Irvine, USA, 
\email{paul.grabowski@uci.edu}
}
%
%
\maketitle

\abstract{Warm dense matter systems created in the laboratory are highly dynamical. In such cases electron
dynamics is often needed to accurately simulate the evolution and properties of the system. Large systems
force one to make simple approximations enabling computationally feasibility.
Wave packet molecular dynamics (WPMD) provides a simple framework for simulating time-dependent quantum plasmas. 
Here, this method is reviewed. The different variants of WPMD are shown and compared and their validity is discussed.
}

\section{Introduction}
\label{sec:1}

The creation of warm dense matter in the laboratory
is a dynamic process. A large amount of energy
is delivered to a target in a short period of time. Several different methods for
delivering this energy have been developed, including exploding wire \cite{DeSilvaKatsouros1998,BenageEtAl1999,DeSilvaVunni2011},
laser foil \cite{MaloneEtAl1975,RipenEtAl1975,MeadEtAl1976,YaakobiBristow1977,AnthesEtAl1978,DecosteEtAl1979,NgEtAl1986,WidmannEtAl2004,PingEtAl2006,PingEtAl2010,ZastrauEtAl2010,ZastrauEtAl2011,RusEtAl2011,InogamovEtAl2011,LiEtAl2012,OgitsuEtAl2012},
inertial confinement fusion \cite{NuckollsEtAl1972,ClarkeEtAl1973},
ion beam \cite{FlippoEtAl2008,HoffmannEtAl2009,LevyEtAl2009,BieniosekEtAl2010,ZhaoEtAl2012,GauthierEtAl2013}, 
Z machine \cite{KnudsonEtAl2003,KnudsonEtAl2004,KnudsonDesjarlais2009,RootEtAl2013},
free electron laser \cite{BostedtEtAl2009,Altarelli2010,Altarelli2011,HauRiege2013},
and laser induced shock \cite{VanKesselSigel1974,VeeserSolem1978,TrainorEtAl1979,CottetEtAl1984,NgEtAl1985,KuglandEtAl2009} experiments. Understanding energy exchange processes is crucial to the proper interpretation
of these experiments. Lasers and ion beams deposit most of their energy in the electrons while
shocks impart most of their energy to the ions. Energy absorption by the electrons in the former cases
leads to a time-dependent light induced ionization and scattering \cite {Pabst2013,ZiajaEtAl2013} or a stopping power problem 
\cite{HoffmanEtAl2009,GrazianiEtAl2011},
respectively, while all of these cases produce electron-ion temperature 
relaxation \cite{BenedictEtAl2012}. 
Furthermore, electrons transport their energy via possibly non-equilibrium 
electrical and thermal conduction.
Recent efforts \cite{Pabst2013,ZiajaEtAl2013} 
to measure electron and ion dynamics at attosecond and picosecond timescales
reinforce the need for a short-timescale simulation capability. It is now possible to
directly image electron density \cite{DixitEtAl2013} and such efforts can validate 
any theoretical predictions. 

Ideally, one should
calculate a numerically converged solution to the many-body time-dependent Schr\"{o}dinger equation for the
electron wave function in the external potential due to the time-dependent electron-ion interactions.
Tens or a few hundred degrees of freedom can be evolved with the efficient multi-configuration
time-dependent Hartree (MCTDH) method \cite{Wang.2003.1289,Mant.2008.164116}. 
Big simulations are needed to solve non-equilibrium systems for statistical reasons, to resolve gradients, and to resolve mean free paths 
of the particles in the system.
Despite making significant progress in reducing computational
effort for many-body quantum problems, MCTDH still scales exponentially with the number of 
degrees of freedom and cannot handle the system sizes needed for these systems.
The obvious simplification would be to use time-dependent density functional theory 
\cite{RungeGross1984} for which one would only need to evolve a three-dimensional density. 
However, very little is known about the accuracy of commonly used
functionals for non-equilibrium high energy density systems.
Finite computational resources then necessitate simple models that capture only the 
requisite quantum mechanics. 

Wave packet molecular dynamics (WPMD) is a simple time-dependent quantum mechanical method with a rich underlying mathematical structure \cite{Littlejohn1986}.
Here, WPMD is defined as the joint propagation of classical ions, represented as point particles,
and quantum
electrons, represented as wave packets. The wave packets are single electron states, localized in both position and momentum,
and are combined via a Hartree product or Slater determinant to form the many-body wave function.
Several groups have used WPMD to calculate a diverse set of observables, including
equations of state \cite{KnaupEtAl2002,KnaupEtAl2003,JakobEtAl2007,SuGoddard2007,JakobEtAl2009,SuGoddard2009}, the collision rate \cite{MorozovValuev2009}, electrical conductivity \cite{JakobEtAl2007,JakobEtAl2009}, the diffusion coefficient \cite{SingerSmith1986}, the dynamic
structure factor \cite{ZwicknagelPschiwul2006}, stopping power \cite{KlakowEtAl1996,ZwicknagelEtAl1999}, and shock Hugoniot curves \cite{KnaupEtAl2003,SuGoddard2007,JakobEtAl2009,KnaupEtAl2001b,Jaramillo-BoteroEtAl2011}.
In this chapter, the different derivations of the equations of motion for the wave packets are summarized in Sec. \ref{sec:2}.
The many different variations in wave packet forms used, antisymmetrization approximations, and interpretation of
variational parameters are reviewed in Sec. \ref{sec:3} and an outlook is given in Sec. \ref{sec:4}.

\section{Theoretical Basis}
\label{sec:2}

\subsection{Ehrenfest's Theorem}

Quantum mechanics is infinitely more complicated than classical mechanics. Quantum dynamics happens
within the full Hilbert space of the system while classical dynamics are described by a path through a finite-dimensional
phase space. The quantum mechanical state does not correspond to any classical quantity such as position or momentum.
We can, however, describe this state by its expectation values of moments of these quantities. Ehrenfest \cite{Ehrenfest1927}
was the first to show how these moments evolve with his famous theorem,
\begin{equation}
\frac{d}{dt}\left\langle\hat{A}\right\rangle=\frac{1}{i\hbar}\left\langle\left[\hat{A},\hat{H}\right]\right\rangle
+\left\langle\frac{\partial\hat{A}}{\partial t}\right\rangle,
\end{equation}
where $\hat{A}$ is an arbitrary operator and $\hat{H}$ is the Hamiltonian operator,
typically given by 
\begin{equation}
\hat{H}=\sum_i \frac{\hat{p}_i^2}{2m_i}+\sum_{i<j}V_{ij}(\hat{\mathbf{r}}_i,\hat{\mathbf{r}}_j).
\end{equation}
Here, the summations are over the particles in the system, $V_{ij}$ is the interaction potential energy between particles $i$ and $j$, and
$\hat{\mathbf{p}}_i$, $\hat{\mathbf{r}}_i$ , and $m_i$ are the momentum operator,
position operator, and mass of particle $i$, respectively.
It is easy to show that
\begin{eqnarray}
\label{Ehrenfest1}\frac{d}{dt}\left\langle\hat{\mathbf{r}}_i\right\rangle& =&\frac{\left\langle\hat{\mathbf{p}}_i\right\rangle}{m_i}\\
\label{Ehrenfest2}\frac{d}{dt}\left\langle\hat{\mathbf{p}}_i\right\rangle& =&-\sum_j\left\langle\mathbf{\nabla}_iV_{ij}\right\rangle\\
\label{Ehrenfest3}\frac{d}{dt}\left\langle\hat{r}_i^2\right\rangle& =&\frac{\left\langle\left\{\hat{\mathbf{r}}_i,\hat{\mathbf{p}}_i\right\}\right\rangle}{m_i}\\
\label{Ehrenfest4}\frac{d}{dt}\left\langle\hat{p}_i^2\right\rangle& =&\frac{1}{i\hbar}\sum_j\left\langle[\hat{p}_i^2,V_{ij}]\right\rangle,
\end{eqnarray}
where $\mathbf{\nabla}_i$ is the gradient with respect to $\mathbf{r}_i$,
the anti-commutator is defined as $\left\{\hat{\mathbf{r}}_i,\hat{\mathbf{p}}_i\right\}=\hat{\mathbf{r}}_i\cdot\hat{\mathbf{p}}_i+\hat{\mathbf{p}}_i\cdot\hat{\mathbf{r}}_i$, and we have suppressed the dependence of $V_{ij}$
on the position operators. These equations are only a few of the infinite number of equations needed to describe complete
quantum dynamics. If $V_{ij}$ is a smooth $C^\infty$ function, then all of the time derivatives 
of $\langle\hat{\mathbf{r}}_i^n\hat{\mathbf{p}}_j^m\rangle$ for any $i,j,n,$ and $m$ will be a function of these same moments.
So Ehrenfest's theorem leads to an infinite hierarchy of equations. These equations can be truncated at finite values of 
$n$ and $m$ with the help of a closure. For example, a common choice of restricted wave function in WPMD is 
the isotropic Gaussian for each quantum particle,
\begin{equation}
\varphi_G(\mathbf{r}_i)=\left(\frac{3}{2\pi\sigma^2}\right)^{3/4}\exp\left[-\left(\frac{3}{4\sigma^2}-\frac{ip_\sigma}{2\hbar\sigma}\right)|\mathbf{r}-\mathbf{r}_i|^2+\frac{i\mathbf{p}\cdot(\mathbf{r}-\mathbf{r}_i)}{\hbar}\right],\label{GaussianWPAnsatz}
\end{equation}
where $\mathbf{r}_i$ is the position coordinate, $\mathbf{r}=\langle\hat{\mathbf{r}}_i\rangle$ is its expectation value, $\mathbf{p}=\langle\hat{\mathbf{p}}_i\rangle$ is the expectation value of momentum, $\sigma = \sqrt{\langle \hat{r}_i^2\rangle-|\langle \hat{\mathbf{r}}_i\rangle|^2}$ is the uncertainty in position, and $p_\sigma$ is its conjugate momentum. Within the Hartree approximation, equations (\ref{Ehrenfest1}-\ref{Ehrenfest4})
imply
\begin{eqnarray}
\dot{\mathbf{r}}&=&\frac{\mathbf{p}}{m_i}\\
\dot{\mathbf{p}}&=&-\sum_j\left\langle\mathbf{\nabla}_iV_{ij}\right\rangle\\
\dot{\sigma}&=&\frac{p_\sigma}{m_i}\\
\dot{p}_\sigma&=&\frac{9\hbar^2}{4m_i\sigma^2}-\frac{1}{\sigma}\sum_j\left\langle(\hat{\mathbf{r}}_i-\mathbf{r})\cdot\mathbf{\nabla}_iV_{ij}\right\rangle.
\end{eqnarray}
These equations of motion predict that the center of the wave packet moves at the expectation value of velocity.
This velocity changes according to the expectation value of the force. The width of the wave packet changes at a rate
of $p_\sigma/m_i$, and its conjugate momentum evolves so as to satisfy the Heisenberg uncertainty principal and
minimize the expectation value of the potential energy.

\subsection{Local Harmonic Approximation}
\label{subsec:2}

Heller's original formulation \cite{Heller1975} of time-dependent semiclassical wave packet dynamics approached the problem
differently. Instead of making a wave function ansatz, he expanded the potential energy to second order in the distance to the
expectation value of the configuration of the system. 
\begin{equation}
\label{VTaylorExp}
V(\mathbf{X})\approx V(\tilde{\mathbf{X}})+\mathbf{\nabla_X}V(\tilde{\mathbf{X}})\cdot(\mathbf{X}-\tilde{\mathbf{X}})+\frac{1}{2}\mathbf{\nabla_X}\otimes\mathbf{\nabla_X}V(\tilde{\mathbf{X}}):(\mathbf{X}-\tilde{\mathbf{X}})\otimes(\mathbf{X}-\tilde{\mathbf{X}}),\label{PotentialExpansion}
\end{equation}
where $\mathbf{X}=\{\mathbf{x}_1,\ldots,\mathbf{x}_N\}$ is the set of positions of all $N$ particles in the system, 
$\tilde{\mathbf{X}}=\langle\mathbf{X}\rangle$ is its expectation value, $\otimes$ is the tensor product, and $:$ indicates the 
contraction of the indices of the rank-two tensors on either side of it. The system is then a $3N$ dimensional harmonic oscillator
with characteristic frequencies that depend directly on $\tilde{\mathbf{X}}$ and so indirectly on time.
The further simplification of neglecting the inter-particle terms in the last term of Eq. \ref{VTaylorExp} and the particle statistics leads
to an anisotropic formulation of WPMD. The potential energy has the reduced form:
\begin{equation}
V(\mathbf{X})\approx \sum_i^N 
V_i(\tilde{\mathbf{x}}_i)
+\mathbf{\nabla}_{\mathbf{x}_i}V(\tilde{\mathbf{x}_i})\cdot(\mathbf{x}_i-\tilde{\mathbf{x}}_i)
+\frac{1}{2}\mathbf{\nabla}_{\mathbf{x}_i}
\otimes\mathbf{\nabla}_{\mathbf{x}_i}
V(\tilde{\mathbf{x}}_i):(\mathbf{x}_i-\tilde{\mathbf{x}}_i)\otimes(\mathbf{x}_i-\tilde{\mathbf{x}}_i),
\end{equation}
where $\tilde{\mathbf{x}}_i=\langle\mathbf{x}_i\rangle$, and the approximate many-body wave function is 
\begin{equation}
\psi(\mathbf{X},t)\approx \prod_i^N \left(\frac{\textnormal{det} \Sigma_i}{\pi}\right)^{1/4}e^{-(\mathbf{x}_i-\mathbf{r}_i)^T\cdot
(\Sigma_i+\textnormal{i}\Pi_i)\cdot(\mathbf{x}_i-\mathbf{r}_i)+\textnormal{i}\mathbf{p}_i\cdot
(\mathbf{x}_i-\mathbf{r}_i)/\hbar+\textnormal{i}\xi_i},
\end{equation}
where $\Sigma_i$ and $\Pi_i$, $\mathbf{r}_i$ and $\mathbf{p}_i$, and $\xi_i$ are time-dependent tensors, vectors, and scalars, 
respectively, which depend on the local value of potential energy and its derivatives.
The main issue in applying this formulation to plasma physics is that
the Coulomb potential is singular. So the expansion (\ref{PotentialExpansion}), is divergent near the singularities.

\subsection{Time Dependent Variational Principle}

One variational principle, proposed by McLachlan \cite{McLachlan1964} and used in Refs. \cite{SingerSmith1986,CorbinSinger1982},
is to minimize 
\begin{equation}
I(\psi,\theta)=\int|i\hbar\theta-\hat{H}\psi|^{2}dV
\end{equation}
 with respect to $\theta=\partial\psi/\partial t$, where the integration
is performed over all of configuration space.

The Dirac-Frenkel time-dependent variational principle (TDVP) \cite{Dirac1930} leads to a rigorous approximation of the time-dependent Schr\"{o}dinger equation
(TDSE) with a given variational ansatz. With this method
the residual of the TDSE is minimized over a given subspace of states
$\left|\psi\right\rangle $, 
\begin{equation}
\delta\int\limits _{t_{i}}^{t_{f}}\left\langle \psi\left|i\hbar\frac{\partial}{\partial t}-\hat{H}\right|\psi\right\rangle dt=0,\label{TDVP}
\end{equation}
where $t_{i}$ and $t_{f}$ are the initial and final times of the
integration, and $\hat{H}$ is the Hamiltonian. If the state
$\left|\psi\right\rangle $ is allowed to vary throughout a Hilbert
space that includes the solution, the TDSE will be exactly solved.
Otherwise, the error in the state grows linearly with time over short
times \cite{FeldmeierSchnack2000}.
This variational principle is equivalent
to the McLachlan approach for the most general form of a Gaussian
wave packet \cite{BroeckhoveEtAl1988}.

A variational state $|\bd{q}\rangle$ can be parametrized by a
vector of complex time-dependent variational parameters, 
\begin{equation}
{\bf q}  =\{{q}_{1},{q}_{2},\ldots,{q}_{N_{v}}\}.
\end{equation}
The variational parameters follow the equations of motion \cite{FeldmeierSchnack2000}:
\begin{equation}
\textnormal{i}\bd{N}\dot{{\bf q}}  =\frac{\partial\langle\hat{H}\rangle}{\partial{\bf q}^{*}},\blue{\quad 
-\textnormal{i}\bd{N}\dot{{\bf q}}^*  =\frac{\partial\langle\hat{H}\rangle}{\partial{\bf q}},}\label{EqMotion}
\end{equation}
where $\langle \hat{H}\rangle=\langle \psi|\hat{H}|\psi\rangle $  and $^{*}$ denotes the complex
conjugate. The Hermitian norm matrix is defined by \cite{FeldmeierSchnack2000}:
\begin{equation}
N_{ab}=\frac{\partial}{\partial q_{a}^{*}}\frac{\partial}{\partial q_{b}}\ln\langle{\bf q}|{\bf q}\rangle.\label{NormMatr}
\end{equation}
\blue{Note, Eqs. (\ref{EqMotion}) are time reversed forms of each other; so models derived from the TDVP preserve time reversal
symmetry.}
For special choices of the variational form and parameters, the matrix
$\bd{N}$ reduces to a \red{{simple diagonal}}\blue{trivially-inverted} matrix and canonical positions and
momenta can be defined that make the equations of motion have a 
Hamilton form in $N_{v}$ dimensions (see for example
Ref.\
\cite{KermanKoonin1976}): 
\begin{equation}
\red{\dot{\bd{r}}} \red{ =\frac{\partial\langle\hat{H}\rangle}{\partial\bd{p}}, }\blue{\dot{\bd{\rho}}} \blue{ =\frac{\partial\langle\hat{H}\rangle}{\partial\bd{\pi}}, }
\quad\red{ \quad\dot{\bd{p}} }\red{ =-\frac{\partial\langle\hat{H}\rangle}{\partial\bd{r}}.}
 \blue{\quad\dot{\bd{\pi}} }\blue{ =-\frac{\partial\langle\hat{H}\rangle}{\partial\bd{\rho}}.}
\label{eq:TDVPmotion}
\end{equation}
In spite of the persuasiveness of this form, it has to be noted that
\red{${\bf r}$}\blue{${\bd{\rho}}$} and \red{${\bf p}$}\blue{${\bd{\pi}}$} are variational parameters inextricably tied to
a particular variational wave function that should not be mistaken for classical
positions and momenta if the quantum nature of the method is to be preserved. Using the TDVP with a small number of parameters
requires physical intuition as to the form of the wave function. It
must be flexible enough to give reasonable observables as well as
numerically convenient and capable of representing the desired initial
state.

%


\section{Usage}
\label{sec:3}

\subsection{Alternate Wave Packet Forms}

The standard isotropic Gaussian form [Eq. (\ref{GaussianWPAnsatz})] for the wave packets in WPMD was mainly chosen for
its mathematical simplicity. Expectation values of the kinetic energy and the potential energy for many different types of interactions
(including the Coulomb potential) can be analytically calculated. This becomes a great utility in many-body dynamics, for which 
extensive calculations limit computationally-feasible system sizes. However, there is much to be desired physically from a variational form that this simple wave packet lacks. Well known asymptotic behaviors both near and far from nuclei are incorrectly modeled 
by Gaussians. The wave packets are unable to breakup and properly share their density with all the nuclei to produce an
accurate representation of a free state, nor can an isotropic Gaussian accurately form bonds, but simple bonding can occur \cite{SuGoddard2009}. Furthermore, there is the
issue of reconciling a periodic system with an aperiodic wave function. Various authors have made attempts to improve all of 
these shortcomings. Their trial wave functions are listed below:

\begin{itemize}
\item{Anisotropic Gaussian \cite{Littlejohn1986}

\begin{equation}
\varphi_{ag}(\x,t) = \left(\frac{\det \Sigma}{\pi}\right)^{1/4}e^{-(\x-\r)^T\cdot\left(\Sigma+i\Pi\right)\cdot(\x-\r)+i\p\cdot(\x-\r)/\hbar+i\xi},\label{AnGaussWP}
\end{equation}
where $\Sigma$ and $\Pi$ are real symmetric matrices and $\Sigma$ is positive definite. This form allows the Gaussians to 
evolve differently in each dimension. It is most useful for systems which have electron densities around each ion that are not 
isotropic, that is, when bonds are important.
}

\item{Hermite Gaussian Wave Packet \cite{LengletMaynard2007}

\begin{equation}
\varphi_{hg}(\x,t) = \sum_{i,j,k\in A} c_{ijk}\varphi_{hg}^{ijk}(\x,t),
\end{equation}
where the Hermite-Gauss functions $\varphi_{hg}^{ijk}$ are 
\begin{equation}
\varphi_{hg}^{ijk}=\prod_{j=x,y,z}h_{n_j}(\sqrt{\omega_j}(r_j-x_j))\exp[-(\omega_j/2+\textnormal{i}\gamma_j)(r_j-x_j)^2+\textnormal{i}p_j(r_j-x_j)],
\end{equation}
the $h_{n_j}$ are the normalized Hermite polynomials of degree $n_j$, $A$ is the set of allowed
triplets, and $\omega_j$, $r_j$, $\gamma_j$, and $p_j$ are variational parameters. The set $A$ 
can be restricted to reduce the numerical effort in inverting the overlap matrix; otherwise, in the limit 
that $A$ includes all possible triplets, the Hermite Gaussian wave packet becomes exact for 
one-electron problems.

}

\item{Split Wave Packet \cite{MorozovValuev2012,GrabowskiEtAl2013}

\begin{equation}
\varphi_s(\bd{x},t)\propto\sum_{\alpha=1}^{M}c_{\alpha}\varphi_{\alpha}(\bd{x},t),\label{WPdef}
\end{equation}
where 
each wave packet ($\varphi_\alpha$) has the same form as Eq. (\ref{GaussianWPAnsatz}),
and the variational parameters ($\bd{r}$, $\bd{p}$, $\sigma$, and $p_\sigma$) take on different values and
evolve independently for each wave packet.  This form represents a single electron wave function
by $M$ Gaussians with mixing coefficients $c_{\alpha}$. It allows the wave packet to breakup, following the behavior
observed in Ref. \cite{GrabowskiEtAl2013}.
}

\item{Periodic Wave Packet \cite{KnaupEtAl2002,KnaupEtAl2003,KnaupEtAl2001b}

\begin{equation}
\varphi_p(\x,t)\propto\sum_\bd{n}\varphi_G(\x-\bd{n}L),
\end{equation}
where $\bd{n}$ is a lattice index for the three-dimensional periodic system and $L$ is the length of the 
periodic box. This form is more consistent with periodic boundary conditions used in bulk plasma simulations. It causes
a large wave packet to be even more weakly interacting with the rest of the system and so be more likely to have divergent
width.
}
\item{Periodic-Bloch Wave Packet \cite{JakobEtAl2007,JakobEtAl2009}

\begin{equation}
\varphi_{pb}(\x,t)=e^{i\bd{q}\cdot\x/\hbar}\varphi_p(\x),
\end{equation}
where $\bd{q}$ is the Bloch momentum.
}

\item{Bound-Free Wave Packets \cite{EbelingMilitzer1997}

\begin{equation}
\varphi_b(\x,t)=\frac{1}{\sqrt{\pi a_0^3}}e^{-|\x-\r_I|/a_0},
\end{equation}
where $a_0$ is the Bohr radius and $\r_I$ is the position of an ion,
\begin{equation}
\varphi_f(\x,t)=\left(\frac{3}{2\pi\sigma^2}\right)^{3/4}e^{3(\x-\r)^2/4\sigma^2+i\p\cdot(\x-\r)/\hbar},
\end{equation}
and the width $\sigma$ is fixed.
}
Ebeling and Militzer used both of these forms to represent bound and free states. The electrons were allowed to transition
between these states according to known cross sections. Of course, the exact bound and free states can differ
significantly from $\varphi_b$ and $\varphi_f$.

\item{Self Similar Wave Packet \cite{MurilloTimmermans2003}

\begin{equation}
\varphi_{ss}(\x,t)=\sqrt{s^{-3}\rho_0\left(\frac{\x-\r}{s}\right)}e^{i[\p\cdot(\x-\r)+p_s(\x-\r)^2/2s]/\hbar+\phi},
\end{equation}
where $s$, $\r$, $\p$ and $p_s$ are time-dependent variational parameters and 
\begin{equation}
\rho_0(\x)=\left\{\begin{array}{ll}
\left(\frac{1}{\pi G^2}\right)^{3/2}e^{-\x^2/G^2}&\textnormal{Gaussian}\\
\frac{1}{8\pi E}e^{-\x/E}&\textnormal{Exponential}
\end{array}\right.
\end{equation}
Murillo and Timmermans compared the relative accuracy of the Gaussian and exponential forms of $\rho_0$
in calculating the ground state energies of hydrogenic systems, helium-like ions, and the hydrogen molecule. 
The exponential wave packet performed better for single atoms, but the Gaussian wave packet better reproduced the hydrogen
molecular bond. It is unclear which is better for time dependent
systems for which $\r$ can become displaced from the positions of the ions or centers of bonds.

}

\item{Harmonically Constrained Wave Packet \cite{EbelingEtAl2006}

\begin{equation}
\varphi_h(\x,t)=\varphi_G(\x,t)e^{id(\x-\r)^4/\sigma^2},\label{HCWP}
\end{equation}
where $d$ is an adjustable parameter which controls how large width $\sigma$ can get. The consequences of fixing $d$
are discussed in Sec. \ref{WidthConstraints}.
}
\end{itemize}

\subsection{Width Constraints\label{WidthConstraints}}

\begin{figure}[b]
\sidecaption
\includegraphics[scale=.65]{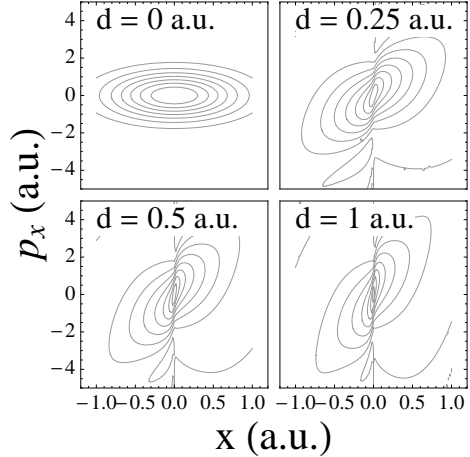}
%
%
\caption{Wigner density cross section of a three-dimensional Gaussian wave function with extra phase factor as in Eq. \ref{HCWP} for different values of $d$. The expectation values of 
position and momentum are zero, the width $\sigma$ is one, and its conjugate momentum $p_\sigma$ is zero.}
\label{WignerEbeling}       
\end{figure}

It has been observed \cite{KnaupEtAl2002,KnaupEtAl2003,MorozovValuev2009,KnaupEtAl2001b,EbelingEtAl2006,
KnaupEtAl1999,KnaupEtAl2000,KnaupEtAl2001a} that at high enough temperatures that the mean width of the wave packets increases without 
bound unless the equations of motion are altered. A diverging width leads to a uniform electron density and diminished electron-ion
and electron-electron interactions, preventing electron equilibration or the correct measurement of transport quantities.

The most common method of preventing width spreading was introduced in Ref. \cite{KnaupEtAl1999} and continued by Refs.
\cite{KnaupEtAl2002,KnaupEtAl2003,MorozovValuev2009,KnaupEtAl2001b,KnaupEtAl2000,KnaupEtAl2001a}. A harmonic
constraint is added to the energy expectation value in an ad hoc fashion,
\begin{equation}
H_{Harm}=a \sigma^2,
\end{equation}
where $a$ is an adjustable parameter setting the mean width of the electrons. Alternatively, Ebeling and coworkers 
\cite{EbelingEtAl2006} were able to derive such an expression by changing the variational wave function to Eq. (\ref{HCWP}),
giving a slightly different harmonic constraint,
\begin{equation}
H_{Harm}'=\frac{20d}{3m}\left(p_\sigma \sigma+\frac{14}{3}d\sigma^2\right).
\end{equation}
Changing the variational wave function to Eq. (\ref{HCWP}) strongly affects  the properties of the wave function. To visualize the
change, the Wigner density
\begin{equation}
f_W(\bd{R},\bd{P},t)=\frac{1}{(2\pi\hbar)^3}\int \varphi_h^*\left(\bd{R}+\frac{\bd{s}}{2}\right)\varphi_h\left(\bd{R}-\frac{\bd{s}}{2}\right)e^{i\bd{P}\cdot \bd{s}/\hbar}d^3s
\end{equation}
was calculated at $\bd{R}=\{x,0,0\}$ and $\bd{P}=\{p_x,0,0\}$ for $\r = 0$, $\p=0$, $p_\sigma = 0$, and $\sigma = 1 a.u.$ for
several different values of $d$ and shown in Fig. \ref{WignerEbeling}. When $d\ne 0$ the Wigner density is twisted and develops
a discontinuity in its derivatives near $x = 0$. 

Morozov and Valuev \cite{MorozovValuev2009} proposed two alternative constraints. They introduced periodic boundary conditions
for the width coordinate (equivalent to a reflecting boundary condition at some maximum width). This change allowed the wave 
packets to equilibrate to a width distribution, but the equilibrated density was too constant. They also invented an energy-based
constraint, in which the wave packets experienced a confining potential only if their interaction potential energy with the nearest 
ion were above a given threshold. By tuning this threshold they could obtain different results for the collision frequency. It is
unlikely that every dynamical quantity will obtain the correct value at the same threshold energy.

\begin{figure}[b]
\sidecaption
\includegraphics[scale=.65]{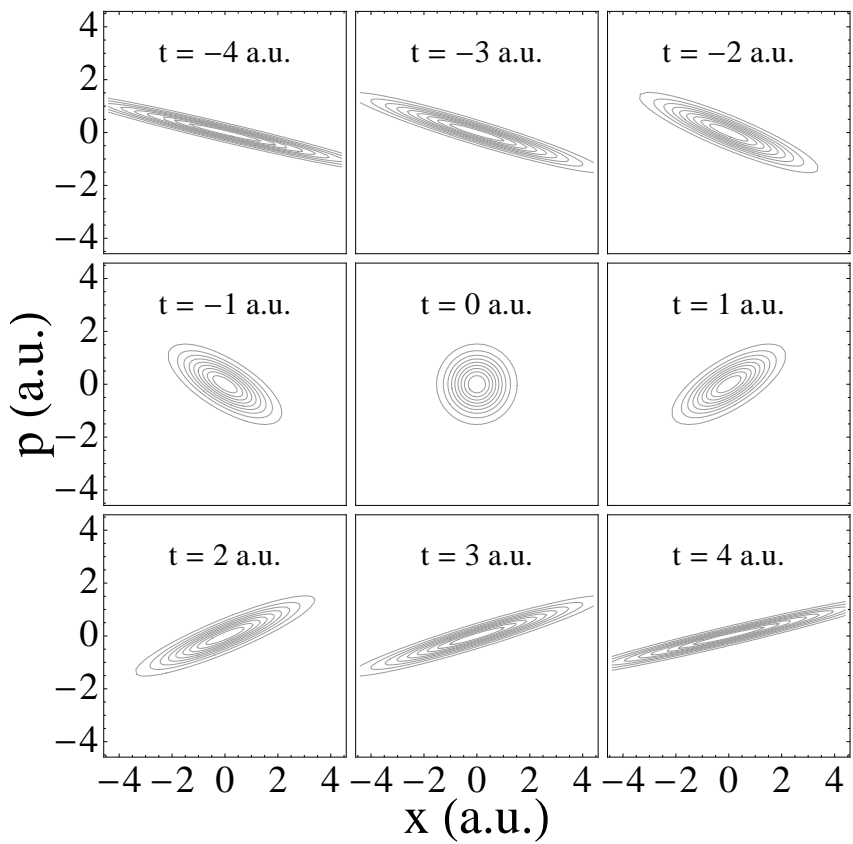}
%
%
\caption{Wigner density of a one-dimensional Gaussian wave function propagating in vacuum. The expectation values of 
position and momentum are zero at all times.}
\label{WignerFreePart}       
\end{figure}

The underlying assumption of the harmonic constraint is that a diverging wave packet width is unphysical. However, it is 
physical for wave packet widths to diverge, but they should localize near ions at the same time. In fact, the simplest
system (a single electron wave packet in a vacuum) has a well known exact solution which exhibits wave packet width divergence.
Isocontours of the Wigner density are shown for the non-interacting wave packet in Fig. \ref{WignerFreePart} at a series
of different times. The time coordinate is defined so that the wave packets reaches a minimum uncertainty configuration at
$t=0$. At negative times the uncertainty in position is decreasing and at positive times it is increasing. The momentum 
uncertainty remains constant for this special case of the non-interacting wave packet because momentum eigenstates are
also energy eigenstates. This behavior is generic to all non-interacting particles. At late times, they will always have a diverging
width and will never reach minimum uncertainty again.

Grabowski {\it et al.} \cite{GrabowskiEtAl2013} showed that this divergence also occurs in their model plasma, made of
a single dynamic electron propagating through a fixed periodic system of statically screened protons. This simple system could 
be propagated exactly and it was observed that the simple wave packet variational form actually underestimates
the spreading, and its main failure was its inability to individually scatter off each proton, which would have created the 
necessary fluctuations in density.

\subsection{Antisymmetrization}

The most computationally expensive part of a wave-function-based quantum calculation is the accurate treatment of Fermi statistics.
Such a treatment requires a fully antisymmetrized many-body wave function, which becomes especially cumbersome when the
single particle states are not orthogonal to each other as in WPMD. The totally antisymmetric wave function is
\begin{equation}
\psi(\x_1,\ldots,\x_N,t)=n\hat{A}\prod_i \varphi_i(\x_i,t)\chi_i, \label{AntisymState}
\end{equation}
where $\{\x_1,\ldots,\x_N\}$ are the coordinates of $N$ electrons, $t$ is time, $\chi_i$ is a Pauli spinor, $n$ is a normalization constant, 
$\varphi_i$ is $i$th wave packet, here of the form Eq. (\ref{GaussianWPAnsatz}), 
$\hat{A}$ is the antisymmetrization operator,
\begin{equation}
\hat{A} = \prod_{i<j}(1 - \hat{\epsilon}_{ij}),
\end{equation}
and $\hat{\epsilon}_{ij}$ is the exchange operator, which exchanges the positions and spins of particles $i$ and $j$. 

\begin{figure}[b]
\includegraphics[scale=.65]{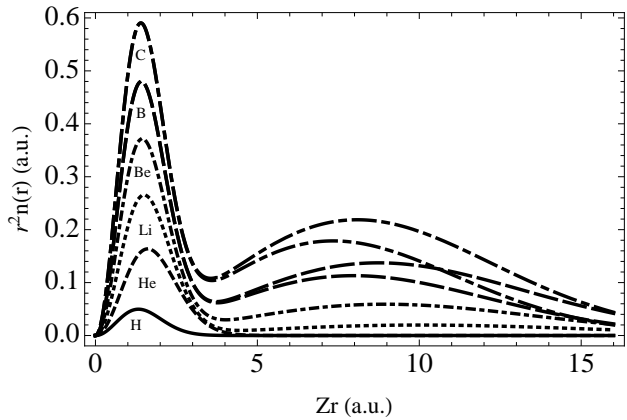}
%
%
\caption{Radial density of fully antisymmetrized variational wave functions for the lowest six elements of the periodic table: hydrogen (solid), helium (dashed), lithium (dotted), beryllium (dot-dashed), boron (long dashed), and carbon (long dot-dashed). Two curves are shown
for boron and carbon to indicate anisotropy. These are the maximum and minimum densities found at each value of $r$.}
\label{RadDens}       
\end{figure}

The expectation value of the Hamiltonian is needed to calculate the equations of motion of the variational parameters 
[see Eq. (\ref{EqMotion})]. The kinetic and potential expectation values for the state (\ref{AntisymState}) are \cite{FeldmeierSchnack2000}
\begin{eqnarray}
\langle\hat{T}\rangle&=&\sum_{k,l=1}^N\langle \varphi_k\chi_k|\hat{t}|\varphi_l\chi_l\rangle\mathcal{O}_{lk}\label{AntisymKinEn}\\
\langle\hat{V}_{ei}\rangle&=&\sum_{k,l=1}^N\langle \varphi_k\chi_k|\hat{v}_{ei}|\varphi_l\chi_l\rangle\mathcal{O}_{lk}\label{AntisymVei}\\
\langle\hat{V}_{ee}\rangle&=&\sum_{k,l,m,n=1}^N\langle\varphi_k\chi_k\varphi_l\chi_l|\hat{v}_{ee}|\varphi_m\chi_m\varphi_n\chi_n\rangle
(\mathcal{O}_{mk}\mathcal{O}_{nl}-\mathcal{O}_{ml}\mathcal{O}_{nk}),\label{AntisymPotEn}
\end{eqnarray}
where $\hat{t}$ is the single-body kinetic energy operator, $\hat{v}_{ei}$ and $\hat{v}_{ee}$ are the two-body electron-ion and
electron-electron potential energy operators, respectively, and $\mathcal{O}_{lk}$
is the inverse of the overlap matrix,
\begin{equation}
(\mathcal{O}^{-1})_{kl}=\langle\varphi_k\chi_k|\varphi_l\chi_l\rangle.
\end{equation} 
It is immediately apparent from Eq. (\ref{AntisymPotEn}) that one must complete an order $N^4$ operation per time step as
well as deal with the possibility that the overlap matrix can become nearly singular, which explains why such
calculations have mainly been limited to systems of finite size, such as the modeling of nucleons in a 
nucleus \cite{FeldmeierSchnack2000}. 

Using Eqs. (\ref{AntisymKinEn}) and (\ref{AntisymPotEn}), the energy and density of atoms can be calculated from the
Raleigh-Ritz variational principle. The electron densities of the first six atoms on the periodic table are shown in Fig. \ref{RadDens}
using individual electron wave packets of the form (\ref{GaussianWPAnsatz}). Even with such a simple variational form,
the correct shell structure is produced. Hydrogen has a single s orbital, helium has two s orbitals of the same size and
opposite spin, lithium has two inner s orbitals and one outer, and beryllium has two inner and two outer  s orbitals. The structure
is more interesting for boron and carbon, in which pairs of wave packets are displaced on opposite sides of the nucleus in 
order to form effective p orbitals. So the boron electron density has a maximum along the $z$ axis and a minimum in the $x$-$y$ plane,
while the carbon atom has maxima along the $x$ and $y$ axes and a minimum on the $z$ axis. Of course, the detailed structure
is wrong; the cusp condition is not satisfied and the density falls off too fast at infinity. However, having shell structure means 
antisymmetrized WPMD includes a simple version of bound-free transitions.

\begin{figure}[b]
\includegraphics[scale=.65]{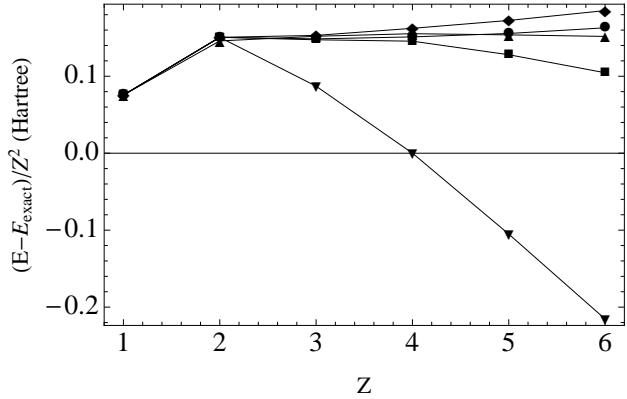}
%
%
\caption{Error in variational energy as a function of Z for different choices of approximation to the energy expectation value: 
Hartree (upside down triangles), antisymmetrized kinetic energy (squares), eFF (triangles), fully antisymmetrized (circles), and 
two-body exchange (diamonds).}
\label{EofZ}       
\end{figure}

Due to the expense of full antisymmetrization, several different approximations have been made in the literature.
The simplest approximation is the Hartree approximation,
\begin{equation}
\hat{A}\approx\hat{A}_1=1,
\end{equation}
used in Refs. \cite{MorozovValuev2009,ZwicknagelPschiwul2006,EbelingEtAl2006}. This approximation is valid at temperatures
much above the Fermi energy. Unfortunately, this is also the regime in which wave packet spreading is the biggest issue, so
width constraints were used in all of those references.

Klakow {\it et al.} \cite{KlakowEtAl1994a,KlakowEtAl1994b} published the first WPMD results for plasmas. They and others 
\cite{KnaupEtAl1999,KnaupEtAl2000,LengletEtAl2006} used the
pairwise antisymmetrization approximation
\begin{equation}
\hat{A}\approx\hat{A}_2=1-\sum_{i<j}\hat{\epsilon}_{ij},
\end{equation}
which includes only two-body exchange. This approximation holds when all of the wave packets are well separated in 
phase space. That is, the distance between their expectation values of position and momentum is big compared to their
uncertainties.

Another approximation \cite{KnaupEtAl2002,KnaupEtAl2003,JakobEtAl2007,JakobEtAl2009,KnaupEtAl2001b} is to use the
Hartree form while calculating the expectation value of the electron-electron potential energy, but use full antisymmetrization for the kinetic energy,
\begin{equation}
\langle V_{ee}\rangle \approx \sum_{k,l=1}^N\langle\varphi_k\chi_k\varphi_l\chi_l|\hat{v}_{ee}|\varphi_k\chi_k\varphi_l\chi_l\rangle.
\end{equation}
Such a scheme allows simulations at temperatures up to $30,000$K before wave packet divergence becomes an issue 
\cite{MorozovValuev2009}.

The simplest computational way of including exchange effects is through the empirical electron Force Field (eFF) model 
\cite{SuGoddard2007,SuGoddard2009,Jaramillo-BoteroEtAl2011}. Energy expectation values are calculated with the 
Hartree approximation, but then corrected with a Pauli potential,
\begin{eqnarray}
E_{Pauli}&=&\sum_{m_{s,i}=m_{s,j}}E(\uparrow\uparrow)_{ij}+\sum_{m_{s,i}\ne m_{s,j}}E(\uparrow\downarrow)_{ij},\\
E(\uparrow\uparrow)_{ij}&=&\left(\frac{S_{ij}^2}{1-S_{ij}^2}+(1-\rho)\frac{S_{ij}^2}{1+S_{ij}^2}\right)\Delta T_{ij},\\
E(\uparrow\uparrow)_{ij}&=&\frac{\rho S_{ij}^2}{1+S_{ij}^2}\Delta T_{ij},\\
\Delta T_{ij} &=& \frac{3}{2}(\bar{\sigma}_i^{-2}+\bar{\sigma}_j^{-2})-\frac{2[3(\bar{\sigma}_i^2+\bar{\sigma}_j^2)-2\bar{r}_{ij}^2]}
{(\bar{\sigma}_i^2+\bar{\sigma}_j^2)^2},\\
S_{ij}&=&\left(\frac{2}{\bar{\sigma}_i/\bar{\sigma}_j+\bar{\sigma}_j/\bar{\sigma}_i}\right)^{3/2}\exp\left(-\frac{\bar{r}_{ij}^2}{\bar{\sigma}_i^2+\bar{\sigma}_j^2}\right),
\end{eqnarray}
where $\bar{\sigma}_i=c_1\sigma_i$, $\bar{r}_{ij}=c_2|\r_i-\r_j|$, and the empirical parameters $\rho = -0.2$, $c_1 = 0.9$ and
$c_2=1.125$ are set by fitting to accurate molecular properties. The form of $E_{Pauli}$ is motivated by terms which appear 
with two-body exchange.

Figure \ref{EofZ} shows the error in energy of the lowest six atoms on the periodic table using all of these 
antisymmetrization methods. Full antisymmetrization makes an error roughly equal to $0.15Z^2$ a.u. This error is due to 
the lack of inclusion of electron correlation and from the simple single electron variational form. Since the approximations
made are with respect to the fully antisymmetrized Gaussian wave packet calculation, their accuracy should be determined
by how close they are to that result. In order of accuracy, the approximations are eFF, two-body exchange, antisymmetrized 
kinetic energy, and the Hartree approximation.

\subsection{Interpretation}

The theoretical basis of WPMD is quantum mechanical. However, the variational parameters describing the electrons' wave
function and the energy expectation value are often interpreted as if they are classical quantities. 
The latter quantity is continuous with a lower bound, necessarily greater than the true ground state due to the restriction 
of the Hilbert space to the subspace spanned by the Gaussian wave packets. There is confusion in the literature as to why 
this energy is not quantized for bound systems, which is taken to be a problem of the model \cite{EbelingMilitzer1997}. 
However, intermediate energy expectation values can easily be reached by creating time-dependent states, which are mixtures
of the states with the discrete energy eigenvalues, which is exactly the case here.

Another confusion arises with respect to the partition function. If the energy expectation value is taken to be classical and the
width taken to be a classical fourth degree of freedom for each electron, then the classical partition function is divergent
unless a width constraint is added. However, this is caused by the approximation
\begin{equation}
e^{-\beta\hat{H}}\approx e^{-\beta\langle\hat{H}\rangle},
\end{equation}
where $\beta$ is the inverse temperature. In fact, exact thermodynamic properties can be calculated in any basis, including
Gaussians. The interested reader is referred to Refs. \cite{FeldmeierSchnack2000} and \cite{MilitzerPollock2000} for accurate
quantum treatments.

Within the plasma physics community dynamical quantities and transport coefficients have been calculated by interpreting the
position and momentum expectation values as classical quantities and then using classical formulas \cite{JakobEtAl2007,JakobEtAl2009,MorozovValuev2009,ZwicknagelPschiwul2006}. This choice along with the width constraints
turns the WPMD model into an effective classical system meant to mimic properties of quantum mechanics; it is no longer
an ab initio quantum model. Tuning this classical model to known quantities may still lead to sensical results.

\section{Outlook}
\label{sec:4}

The dream of having a simple and accurate time-dependent method capable of accurately simulating WDM systems has not been
fully realized. To model high temperature systems, of order or greater than the Fermi energy, either an unphysical width constraint
must be used or the system develops an unphysical homogeneous density. At lower temperatures, the eFF variant of WPMD
has made the greatest strides having been applied to such diverse calculations such as stopping power \cite{SuGoddard2009},
equation of state \cite{SuGoddard2007}, lithium cluster-slab impact \cite{Jaramillo-BoteroEtAl2011}, and the shock Hugoniot 
curves of liquid hydrogen and deuterium \cite{SuGoddard2007,SuGoddard2009,Jaramillo-BoteroEtAl2011}. However, 
the empirical parameters in this model make it difficult to quantify errors a priori. 

The greatest errors in WPMD stem from its simple variational form. Improvements for high temperature systems should focus
on more accurate representations of delocalized free states that have higher densities near ions. The drive for a better representation
of low temperature systems should focus on improving the ability of the wave packets to form bonds by allowing anisotropies.

\begin{acknowledgement}
The author would like to thank Michael S. Murillo for mentorship while he was learning about WPMD and writing this review, John Benage for helpful discussions and background information,
Ronald Redmer for useful suggestions on improving the content of this review,
and Frank Graziani, the organizing committee, and the staff of the Institute for Pure and Applied Mathematics for organizing the workshop: Computational Challenges in Warm Dense Matter.

This work was mostly written during the time the author was an employee of the
Los Alamos National Security, LLC.\ (LANS), operator of the Los Alamos
National Laboratory under Contract No.\ DE-AC52-06NA25396 with the
U.S.\ Department of Energy and funded by the Laboratory
Directed Research and Development Program at LLNL under project tracking
code 09-SI-011.
\end{acknowledgement}

%


\bibliographystyle{spphys}
\bibliography{WPMD.bib,WPMDnSOFT.bib}
\end{document}